\documentclass[aps,prb,showpacs,twocolumn]{revtex4-1}
\usepackage{amsfonts}
\usepackage{amssymb}
\usepackage{amsmath}
\usepackage{graphicx}
\usepackage{epsfig}

\begin{document}

\title{Emergent SSH physics and localization in a cavity--atom system beyond
the rotating-wave approximation}
\author{H. Shen}
\author{J. Cao}
\author{R. Wang}
\email{wangr@tjnu.edu.cn}
\affiliation{College of Physics and Materials Science, Tianjin Normal
University, Tianjin 300387, China}
\author{X. Z. Zhang}
\email{zhangxz@tjnu.edu.cn}
\affiliation{College of Physics and Materials Science, Tianjin Normal
University, Tianjin 300387, China}

\begin{abstract}
We investigate a cavity--atom system beyond the rotating-wave approximation
and recast its dynamics as an effective tight-binding problem in the
excitation-number basis. The same experimentally relevant light--matter
platform exhibits two distinct regimes controlled by the photon number. In
the large-photon-number regime, the hopping amplitudes become nearly uniform
and the effective lattice reduces to a dimerized SSH-type chain subject to a
linear energy gradient. The spectrum then organizes into one or two
Wannier--Stark ladders, giving rise to controllable Bloch and Bloch--Zener
oscillations and quantitatively explaining the observed revival patterns. In
the small-photon-number regime, the intrinsic $\sqrt{\bar n+n}$ dependence
produces a pronounced hopping deformation, namely dimerized couplings that
increase along the excitation-number lattice. This deformation reshapes the
density of states, enhances collective localization, and generates an
energy-resolved spatial bias of the eigenstates. The corresponding real-time
dynamics displays direction-dependent anomalous diffusion together with
Bloch-like revivals and beat phenomena in selected parameter windows. Our
results show that the same cavity--atom platform can realize both emergent
ladder spectra and hopping-induced localization in excitation-number space.
\end{abstract}

\maketitle

\affiliation{College of Physics and Materials Science, Tianjin Normal
University, Tianjin 300387, China}

\section{Introduction}

Cavity and circuit quantum electrodynamics (\textrm{QED}) \cite{Blais2021}
naturally bring together few-body quantum optics \cite%
{JC1963,Raimond2001,Thompson1992,Ritsch2021,Andreas2022,David2022},
controllable light-matter coupling \cite%
{Wallraff2004,Simmonds2014,Solano2019,Nori2024,Alejandro2024}, and synthetic
lattice physics \cite{Celi2014,Goldman2016,Tomoki2019,Segev2021,GGC2022}
within a single experimentally accessible platform. In particular,
superconducting microwave resonators coupled to artificial two-level atoms
\cite{Nori2019,Wallraff2004,Niemczyk2010,Solano2019} offer a flexible route
to the ultrastrong-coupling regime, where the rotating-wave approximation
breaks down and counter-rotating processes participate directly in the
dynamics. In such devices, the cavity field can be resolved in the Fock
basis, the qubit frequency and the light-matter coupling can be tuned in
situ, and time-resolved population dynamics can be monitored with standard
spectroscopic protocols. These capabilities make superconducting-circuit
architectures a natural quantum-simulation platform for the
excitation-number-space transport problem studied below.

With this experimental platform in mind, we consider a single microwave
resonator coupled to an effective two-level artificial atom and analyze its
dynamics beyond the rotating-wave approximation. In the absence of further
engineering, the system reduces to the quantum Rabi model \cite%
{Rabi1937,Braak2011}. More generally, parametrically assisted couplers and
digital-analog simulation protocols \cite%
{XQT2014,Jochen2017,Simmonds2014,YL2017} enable effective anisotropic
couplings between the co-rotating and counter-rotating channels, motivating
the generalized form of the Hamiltonian adopted in this work. Such a
formulation keeps the connection to an experimentally implementable
superconducting-circuit simulator \cite{DiCarlo2017, CWW2019, GGC2021,
XQK2024,SPT2024, ZJJ2025}\ explicit while allowing the two coupling channels
to be tuned independently.

The main conceptual step of this work is to recast the cavity-atom problem
in the excitation-number basis. Upon this mapping, the full quantum dynamics
is viewed as motion on an effective one-dimensional lattice with
parity-resolved sectors, alternating couplings, and photon-number-dependent
hopping amplitudes \cite{Jonas2023}. This picture places the system in
direct contact with lattice transport phenomena while preserving its
microscopic cavity-\textrm{QED} origin, and already reveals that the same
platform supports two qualitatively different regimes depending on the
photon number.

In the large-photon-number regime, the photon-number dependence of the
hopping amplitudes becomes weak, and the effective lattice reduces to a
dimerized \textrm{SSH}-type chain in a linear energy gradient. The spectrum
then organizes into Wannier--Stark ladders supporting Bloch and Bloch-Zener
oscillations. In the small-photon-number regime, by contrast, the hopping
amplitudes retain a strong photon-number dependence. The resulting hopping
deformation reshapes the density of states, enhances localization, and
generates an energy-resolved spatial bias of the eigenstates, so that
transport becomes strongly state dependent and can display
direction-dependent anomalous diffusion \cite{Metzler1999, ZH2006, ZHP2024},
Bloch-like revivals \cite{WR2024,ZHP2024-2,ZHP2025,Jonas2018}, and beat
phenomena \cite{ZG2015,Maksimov2015}. Our goal is therefore not only to
analyze a generalized cavity-atom Hamiltonian, but also to show how a
realistic superconducting-circuit quantum simulator can interpolate between
emergent ladder physics and hopping-induced localization within the same
microscopic setting. By combining spectral analysis, eigenstate diagnostics,
and real-time dynamics, we establish a unified picture of how
counter-rotating light-matter coupling reorganizes excitation-number space
and produces experimentally testable transport signatures.

The remainder of the paper is organised as follows. In Sec. \ref%
{sec:model_hamiltonian}, we introduce the Hamiltonian for a superconducting
resonator coupled to a two-level artificial atom and map it onto an
effective tight-binding problem in the excitation-number basis. Sec. \ref%
{Large_photon_number} is devoted to the large-photon-number regime, in which
we demonstrate that the effective excitation-number lattice realizes a
dimerized \textrm{SSH}-type structure subject to a uniform Wannier--Stark
potential. In Sec. \ref{Dynamics}, we study the quantum dynamics of $H_{%
\mathrm{eff,1}}$ in the excitation-number lattice, where the interplay
between Wannier--Stark localization and \textrm{SSH}-type dimerization
enables flexible control of Bloch and Bloch-Zener oscillations,
quantitatively explaining the observed revival patterns. Sec. \ref%
{small_photon} demonstrates that in the small-photon-number regime, the
effective excitation-number lattice becomes strongly inhomogeneous, and the
clean Bloch-type motion is replaced by state-dependent transport governed by
the detailed structure of the spectrum, the density of states, and the
spatial organization of the eigenstates. We conclude in Sec. \ref{Summary}
with a summary of our findings and a discussion of promising future
directions.

\section{Model Hamiltonian}

\label{sec:model_hamiltonian}

We model the experimentally relevant situation of a superconducting
microwave resonator coupled to an effective two-level artificial atom. In
the ultrastrong-coupling regime, the rotating-wave approximation is not
controlled, and the system is naturally described by a generalized quantum
Rabi Hamiltonian. From the viewpoint of quantum simulation, the same
structure can also be engineered in a parametrically controlled circuit-QED
device, where the co-rotating and counter-rotating channels are tuned
independently at the effective level. We therefore work with the anisotropic
form below, which contains both the physical Rabi point and its
experimentally motivated generalizations.

Beyond the rotating-wave approximation, the light--matter interaction is
described by the generalized quantum Rabi Hamiltonian
\begin{equation}
H=\omega a^{\dagger }a+\frac{1}{2}\Omega \sigma _{z} +\lambda _{1}(\sigma
^{\dagger }a+\sigma ^{-}a^{\dagger }) +\lambda _{2}(\sigma ^{\dagger
}a^{\dagger }+\sigma ^{-}a),  \label{H_org}
\end{equation}
where $a^{\dagger }$ ($a$) is the creation (annihilation) operator of the
cavity mode with frequency $\omega$, and $\Omega$ denotes the atomic
transition frequency. The states $|g\rangle$ and $|e\rangle$ represent the
ground and excited states of the atom, respectively, with $\sigma
_{z}=|e\rangle\langle e|-|g\rangle\langle g|$ and $\sigma ^{\dagger
}=|e\rangle\langle g|$ ($\sigma ^{-}=|g\rangle\langle e|$). Throughout this
work, we set $\hbar =1$.

The first interaction term proportional to $\lambda _{1}$ corresponds to the
energy-conserving light--matter coupling retained under the rotating-wave
approximation, while the second term proportional to $\lambda _{2}$
describes the counter-rotating processes that simultaneously create or
annihilate atomic and photonic excitations. Such counter-rotating terms
become indispensable in the ultrastrong and deep-strong coupling regimes,
which have been realized in a variety of experimental platforms \cite%
{Solano2019, Niemczyk2010, DiCarlo2017}. In these regimes, the breakdown of
excitation-number conservation leads to fundamentally new dynamical and
spectral properties.

Despite the absence of excitation-number conservation, the Hamiltonian (\ref%
{H_org}) preserves a discrete $\mathbb{Z}_{2}$ parity symmetry generated by
the operator
\begin{equation}
\Pi =(-1)^{\mathcal{N}},\text{ }\mathcal{N}=a^{\dagger }a+\frac{1}{2}(\sigma
_{z}+1),
\end{equation}%
which measures the total excitation number modulo two. As a consequence, the
Hilbert space can be decomposed into two invariant subspaces with odd and
even parity, allowing the Hamiltonian to be block diagonalized accordingly.
This parity symmetry plays a central role in structuring the spectrum and
will be essential for the emergent lattice description discussed below.

By projecting Eq.\ (\ref{H_org}) onto the odd-parity subspace, the
Hamiltonian can be mapped onto an effective one-dimensional lattice model,
\begin{eqnarray}
H_{o} &=&\sum_{n=1}^{N}[\lambda _{1}\sqrt{\bar{n}+2n-1}(|2n-1\rangle
_{oo}\langle 2n|)  \notag \\
&&+\lambda _{2}\sqrt{\bar{n}+2n}(|2n\rangle _{oo}\langle 2n+1|)]+\mathrm{H.c.%
}  \notag \\
&&+\sum_{n=1}^{2N+1}[(\bar{n}+n-1)\omega +\frac{\Omega }{2}%
(-1)^{n+1}](|n\rangle _{oo}\langle n|),  \label{Ho}
\end{eqnarray}%
where $\{|n\rangle _{o}\}$ denotes an effective basis obtained from the
odd-parity states $\{|e,\bar{n}\rangle ,|g,\bar{n}+1\rangle ,|e,\bar{n}%
+2\rangle ,\ldots \}$. Here {$\bar{n}$ }denotes the average photon number
around which the excitation manifold is constructed.

Similarly, in the even-parity subspace spanned by $\{|g,\bar{n}\rangle ,|e,%
\bar{n}+1\rangle ,|g,\bar{n}+2\rangle ,\ldots \}$, the effective Hamiltonian
reads
\begin{eqnarray}
H_{e} &=&\sum_{n=1}^{N}[\lambda _{2}\sqrt{\bar{n}+2n-1}(|2n-1\rangle
_{ee}\langle 2n|)  \notag \\
&&+\lambda _{1}\sqrt{\bar{n}+2n}(|2n\rangle _{ee}\langle 2n+1|)]+\mathrm{H.c.%
}  \notag \\
&&+\sum_{n=1}^{2N+1}[(\bar{n}+n-1)\omega +\frac{\Omega }{2}%
(-1)^{n}](|n\rangle _{ee}\langle n|).  \label{He}
\end{eqnarray}

The detailed derivation of Eqs. (\ref{Ho}) and (\ref{He}) is presented in
Appendix \ref{Appendix:A}. As will be shown in the following sections, the
competition between the alternating hopping amplitudes and the staggered
on-site potential, controlled by the ratios among $\omega $, $\Omega $, and
the coupling strengths $\lambda _{1(2)}$, gives rise to rich localization
features in different photon-number regimes.

\section{Large-photon-number regime}

\label{Large_photon_number}

In this section, we analyze the system in the large-photon-number regime ({$%
\bar{n}=10^{3}$}) and demonstrate that the effective excitation-number
lattice realizes a dimerized \textrm{SSH}-type structure subject to a
uniform Wannier--Stark potential. For clarity, we focus on the even-parity
subspace and omit the subspace index $e$ hereafter.

We consider the regime
\begin{equation}
\bar{n}\gg 2N,
\end{equation}%
where $N$ characterizes the spatial extent of the relevant wave packets in
the excitation-number space. In this limit, the hopping amplitudes in Eq. (%
\ref{He}) can be approximated as
\begin{equation}
\sqrt{\bar{n}+2n-1}\simeq \sqrt{\bar{n}+2n}\simeq \sqrt{\bar{n}},
\label{large condi}
\end{equation}%
up to corrections of order $\mathcal{O}(n/\bar{n})$. Introducing the
parametrization
\begin{equation}
\lambda _{2}\sqrt{\bar{n}}=\kappa (1+\Delta ),\text{ }\lambda _{1}\sqrt{\bar{%
n}}=\kappa (1-\Delta ),
\end{equation}%
and neglecting the global energy offset $(\bar{n}+N)\omega $, the effective
Hamiltonian can be rewritten as
\begin{equation}
H_{\mathrm{eff,1}}=H_{\mathrm{stark}}+H_{\Delta }+H_{\Omega },  \label{Heff1}
\end{equation}%
where each term has a clear physical origin. The dominant contribution is
the Wannier--Stark Hamiltonian
\begin{equation}
H_{\mathrm{stark}}=H_{\mathrm{unif}}+H_{\omega },  \label{Hstk}
\end{equation}%
with
\begin{equation}
H_{\mathrm{unif}}=\sum_{n=1}^{2N}(\kappa |n\rangle \langle n+1|+\mathrm{H.c.}%
),
\end{equation}%
and
\begin{equation}
H_{\omega }=\sum_{n=1}^{2N+1}[n-(N+1)]\omega |n\rangle \langle n|.
\end{equation}%
The linear potential $H_{\omega }$ originates from the bare cavity energy $%
\omega a^{\dagger }a$ and generates a uniform energy gradient in the
excitation-number lattice, leading to Wannier--Stark localization. In
addition, the alternating hopping term
\begin{equation}
H_{\Delta }=\sum_{n=1}^{N}\kappa \Delta \left( |2n-1\rangle \langle
2n|-|2n\rangle \langle 2n+1|+\mathrm{H.c.}\right) ,  \label{H1}
\end{equation}%
introduces SSH-type dimerization, while the staggered on-site potential
\begin{equation}
H_{\Omega }=\sum_{n=1}^{2N+1}\frac{\Omega }{2}(-1)^{n}|n\rangle \langle n|
\label{H2}
\end{equation}%
originates from the atomic degree of freedom and further couples the two
sublattices.

We first consider the unperturbed Wannier--Stark problem,
\begin{equation}
H_{\mathrm{stark}}|\varphi _{m}\rangle =\epsilon _{m}|\varphi _{m}\rangle ,
\end{equation}%
the eigenenergies form an equally spaced ladder,
\begin{equation}
\epsilon _{m}=[m-(N+1)]\omega ,
\end{equation}%
and the corresponding eigenstates are exponentially localized around site $m$%
. In real space, the eigenfunctions can be expressed as
\begin{equation}
|\varphi _{m}\rangle =\sum_{n=1}^{2N+1}J_{n-m}(-\frac{2\kappa }{\omega }%
)|n\rangle ,
\end{equation}%
where $J_{n}$ denotes the Bessel function of the first kind.

Treating $H_{\Delta }+H_{\Omega }$ as perturbations, the first-order
corrected eigenenergies of $H_{\mathrm{eff,1}}$ are given by
\begin{equation}
E_{m}=\epsilon _{m}+(-1)^{m}\gamma ,  \label{Eeff2}
\end{equation}%
with
\begin{equation}
\gamma =\frac{\Omega }{2}J_{0}(-\frac{4\kappa }{\omega })-2\kappa \Delta
J_{1}(-\frac{4\kappa }{\omega }).
\end{equation}%
The first-order corrected eigenfunction of $H_{\mathrm{eff,1}}$ has the form
\begin{eqnarray}
\left\vert \Psi _{m}\right\rangle &=&\left\vert \varphi _{m}\right\rangle
+(-1)^{m}\frac{1}{2\omega }\{2\kappa \Delta \lbrack J_{0}(-\frac{4\kappa }{%
\omega })-J_{2}(-\frac{4\kappa }{\omega })]  \notag \\
&&+\Omega J_{1}(-\frac{4\kappa }{\omega })\}(|\varphi _{m+1}\rangle
+|\varphi _{m-1}\rangle ).  \label{Phieff2}
\end{eqnarray}%
The corresponding eigenstates acquire weak hybridization between neighboring
Wannier--Stark states. Further details of the perturbation expansion
derivation are given in Appendix \ref{Appendix:B}.

Equation\ (\ref{Eeff2}) reveals that the original single Wannier--Stark
ladder splits into two interlaced ladders distinguished by the parity of $m$%
. Within the parameter window $\gamma \in \lbrack 0,\omega )$, three
representative cases arise:

(i) $\gamma =0$: the spectrum retains a single equally spaced ladder with
the energy spacing between adjacent levels $\mathcal{E}=\omega $.

(ii) $\gamma =\omega /2$: the spectrum exhibits extensive double degeneracy
and an emergent ladder with spacing $\mathcal{E}=2\omega $, leading to
frequency-doubled dynamics.

(iii) $\gamma \neq 0$, and $\gamma \neq \omega /2$: the spectrum consists of
two offset ladders with distinct spacings $\mathcal{E}^{(1)}=\omega _{1}$
and $\mathcal{E}^{(2)}=\omega _{2}$ satisfying $\mathcal{E}^{(1)}+\mathcal{E}%
^{(2)}=2\omega $.

The three spectral structures are schematically illustrated in Fig.\ \ref%
{fig1} and will be shown to give rise to qualitatively different Bloch and
Bloch--Zener dynamics in the following subsection.
\begin{figure}[tbp]
\centering
\includegraphics[bb=90 57 685 425, width=9.5cm, clip]{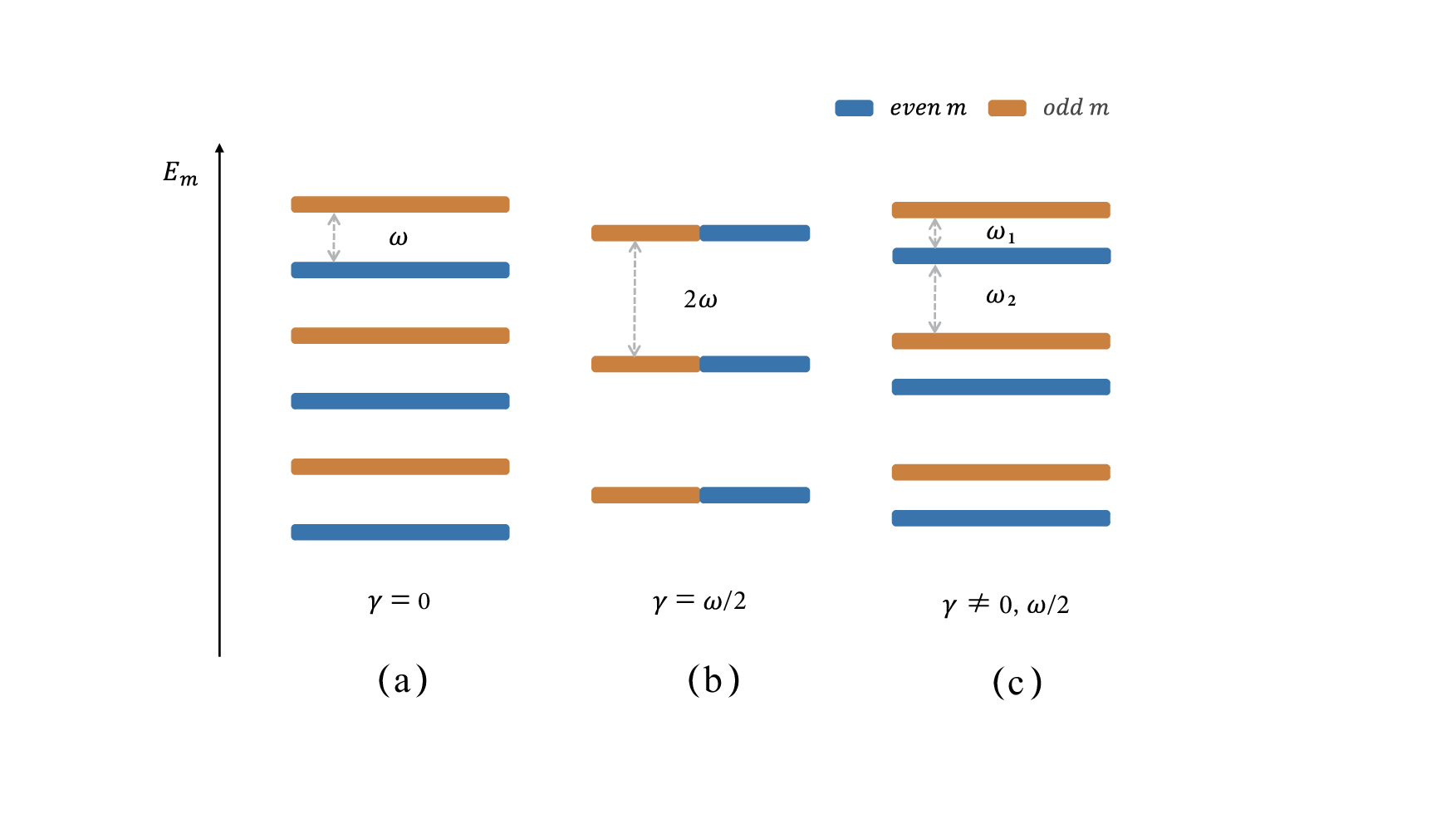}
\caption{Schematic illustration of the eigenenergy structure $E_{m}$ given
by Eq.~(\protect\ref{Eeff2}). Blue (yellow) solid lines denote energy levels
with even (odd) index $m$. (a) $\protect\gamma =0$: all eigenenergies form a
single Wannier--Stark ladder with a uniform level spacing $\protect\omega$.
(b) $\protect\gamma =\protect\omega /2$: the spectrum becomes pairwise
degenerate, and the energy levels reorganize into an effective single ladder
with doubled spacing $2\protect\omega $. (c) $\protect\gamma \neq 0$ and $%
\protect\gamma \neq \protect\omega /2$: the degeneracy is lifted and the
spectra split into two interlaced Wannier--Stark ladders with distinct
spacings $\protect\omega _{1} $ and $\protect\omega _{2}$, satisfying $%
\protect\omega _{1}+\protect\omega _{2}=2\protect\omega $. The three typical
spectral configurations underlie the emergence of single-frequency Bloch
oscillations, frequency-doubled Bloch oscillations, and Bloch--Zener
oscillations, respectively, as discussed in the subsequent dynamical
analysis.}
\label{fig1}
\end{figure}
\begin{figure*}[tbp]
\centering
\includegraphics[bb=61 90 1587 582, width=18cm, clip]{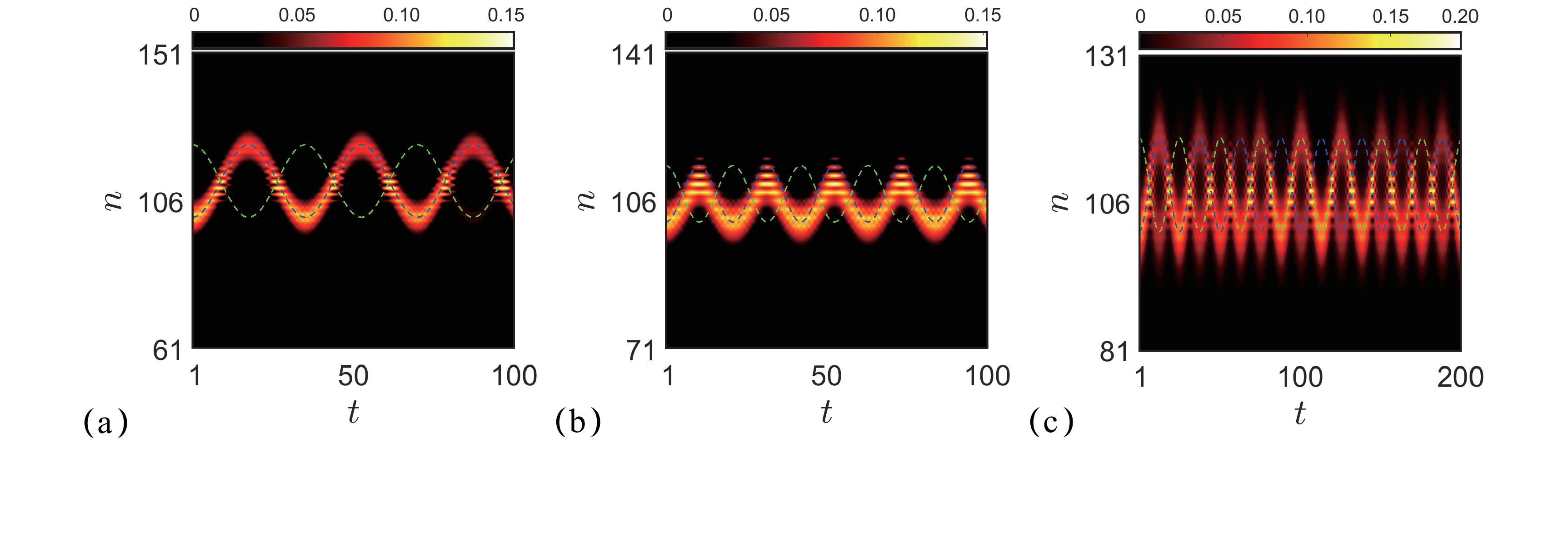}
\caption{Time-evolution probability distribution in the coordinate space for
three typical values of the effective coupling $\protect\gamma $. The blue
and green dashed lines indicate the trajectories of the effective
wave-packet centers $n_{a}(t)$ and $n_{b}(t)$, respectively. (a) $\protect%
\gamma =0$: the dynamics exhibits conventional Bloch oscillations with
period $T_{B}=2\protect\pi /\protect\omega $; (b) $\protect\gamma =\protect%
\omega /2$: the frequency-doubled Bloch oscillations with period {$%
T_{B}^{(2)}=T_{B}/2=\protect\pi /\protect\omega $}. (c) $\protect\gamma =%
\protect\omega /50$: the coexistence of two inequivalent energy ladders
induce Bloch--Zener oscillations characterized with period $T_{\mathrm{BZ}}=2%
\protect\pi /|\mathcal{E}^{(1)}-\mathcal{E}^{(2)}|$. The initial state is
chosen as a Gaussian wave packet, $|\Phi \left( 0\right) \rangle =\frac{1}{%
\protect\sqrt{R}}\sum_{n=1}^{2N+1}e^{-\frac{\protect\alpha ^{2}}{2}%
(n-n_{0})^{2}}|n\rangle $, with $n_{0}=101$, $\protect\alpha =0.2$, and $%
R=8.86$. Other parameters are fixed to $\protect\kappa =1$, $N=100$ and (a) $%
\protect\omega =\Omega =\mathcal{E}=0.18$, $\Delta =0.04$; (b) $\protect%
\omega =\Omega =\mathcal{E}/2=0.15$, $\Delta =-0.24$; (c) $\protect\omega %
=\Omega =\mathcal{E}=0.25$, $\Delta =0.15$ with $\mathcal{E}^{(1)}=0.278$
and $\mathcal{E}^{(2)}=0.222$.}
\label{fig2}
\end{figure*}

\section{Quantum dynamics in the large-photon-number regime}

\label{Dynamics}

Based on the effective Hamiltonian $H_{\mathrm{eff,1}}$ derived in the
previous section, we now investigate the quantum dynamics in the
excitation-number lattice. In the large-photon-number limit, the system
admits an effective Wannier--Stark description, where the cavity frequency $%
\omega $ generates a uniform energy gradient, while the counter-rotating
interaction induces an \textrm{SSH}-type dimerization between neighboring
excitation manifolds.

Owing to the parity conservation of the full Hamiltonian, the time evolution
can be conveniently analyzed in the eigenbasis of $H_{\mathrm{eff,1}}$. A
general initial state is written as
\begin{equation}
|\Phi (0)\rangle =\sum_{m}c_{m}|\Psi _{m}\rangle ,  \label{phi_0}
\end{equation}%
where $|\Psi _{m}\rangle $ denotes the eigenstates of $H_{\mathrm{eff,1}}$,
and {$c_{m}$\ represents }the expansion coefficient of the initial state in
the eigenenergy basis, satisfying $\sum_{m}|c_{m}|^{2}=1$. The time-evolved
state is then given by
\begin{equation}
|\Phi (t)\rangle =e^{-iH_{\mathrm{eff,1}}t}|\Phi (0)\rangle .  \label{phi_t}
\end{equation}

As shown in Sec. \ref{Large_photon_number}, the spectrum of {$H_{\mathrm{%
eff,1}}$ }consists of either one or two equally spaced Wannier--Stark
ladders, depending on the effective perturbation parameter $\gamma $.
Consequently, the quantum dynamics is governed by the interference of one or
two characteristic Bloch frequencies, giving rise to distinct oscillatory
behaviors in the excitation-number space.

\subsection{Single-ladder regime}

We first consider the case $\gamma=0$, where the effective spectrum reduces
to a single Wannier--Stark ladder with uniform level spacing $\omega$, $%
E_m=\epsilon_m$. In this regime, the staggered hopping and on-site
modulation cancel at the leading order, and the dynamics is solely dictated
by the uniform energy gradient.

For an initially localized state $|\Phi (0)\rangle $, the wave packet
undergoes undamped Bloch oscillations with a fundamental period
\begin{equation}
T_{B}=\frac{2\pi }{\mathcal{E}}=\frac{2\pi }{\omega }.
\end{equation}%
The probability distribution periodically revives without noticeable
dispersion, reflecting the exact equidistance of the energy spectrum and the
strong Wannier--Stark localization of the eigenstates.

This behavior is illustrated in Fig. \ref{fig2}(a), where the
excitation-number distribution oscillates rigidly in time and returns to its
initial configuration after each Bloch period.

\subsection{Degenerate ladder and frequency-doubled oscillations}

At $\gamma=\omega/2$, the spectrum develops extensive degeneracy:
neighboring Wannier--Stark levels collapse into pairs, effectively
reorganizing the original ladder with spacing $\omega$ into a new ladder
with spacing $2\omega$.

If the initial state occupies a single sublattice sector, the dynamics is
governed by this reconstructed ladder and remains strictly periodic, with a
reduced oscillation period
\begin{equation}
T_{B}^{(2)}=\frac{2\pi }{\mathcal{E}}=\frac{\pi }{\omega }.
\end{equation}%
This frequency-doubled Bloch oscillation is clearly visible in Fig. \ref%
{fig2}(b), where the revival time is halved compared to the single-ladder
case. Physically, the exact degeneracy suppresses inter-ladder transitions,
leading to an effective decoupling between the two Wannier--Stark
sublattices.

\subsection{Two-ladder regime and Bloch--Zener oscillations}

For generic values $\gamma \neq 0$ and $\omega /2$, the spectrum splits into
two interlaced Wannier--Stark ladders with distinct offsets. We define the
corresponding energy spacing between adjacent levels$\ $as $\mathcal{E}%
^{(1)}=E_{2n}-E_{2n-1}$ and $\mathcal{E}^{(2)}=E_{2n+1}-E_{2n}$, which
satisfy $\mathcal{E}^{(1)}+\mathcal{E}^{(2)}=2\omega $.

In this regime, a generic initial state contains components from both
ladders and evolves under the interference of two Bloch frequencies. As a
result, the excitation-number distribution exhibits pronounced beating
patterns known as Bloch--Zener oscillations, as shown in Fig.\ \ref{fig2}%
(c). The wave packet periodically splits and recombines, signaling coherent
tunneling between the two ladders mediated by the SSH-type dimerization. The
minimal revival time is determined by the frequency mismatch,
\begin{equation}
T_{\mathrm{BZ}}=\frac{1}{|1/T_{1}-1/T_{2}|},
\end{equation}%
where $T_{1}=2\pi /\mathcal{E}^{(1)}$, and $T_{2}=2\pi /\mathcal{E}^{(2)}$.

This dynamics admits an intuitive wave-packet interpretation. In the
large-photon-number regime, the time-evolved state can be approximately
expressed as
\begin{equation}
|\Phi (t)\rangle =\cos (\gamma t)|\varphi _{0}\left( t\right) \rangle -i\sin
(\gamma t)|\varphi _{0}(t+T_{B}/2)\rangle ,  \label{phi_t2}
\end{equation}%
where the time evolution of the same initial state $|\Phi (0)\rangle $ under
the Wannier--Stark Hamiltonian$\ H_{\mathrm{stark}}$ can be written as $%
|\varphi _{0}\left( t\right) \rangle =e^{-iH_{\mathrm{stark}}t}|\Phi
(0)\rangle $. Equation (\ref{phi_t2}) reveals that the dynamics is composed
of two Bloch oscillations shifted by half a Bloch period $T_{B}$, whose
relative weights are periodically exchanged at a rate controlled by the
effective coupling $\gamma $. This coherent population transfer between the
two phase-shifted wave packets provides a transparent physical origin of the
Bloch--Zener oscillations observed in the numerical simulations.

To further quantify the oscillatory behavior, we introduce the trajectories
of the centers of the two effective wave packets,
\begin{equation}
n_{a}(t)=n_{0}+\frac{4\kappa }{\mathcal{E}}\sin ^{2}(\frac{\mathcal{E}t}{2}),
\end{equation}%
and
\begin{equation}
n_{b}(t)=n_{0}+\frac{4\kappa }{\mathcal{E}}\cos ^{2}(\frac{\mathcal{E}t}{2}),
\end{equation}%
where $n_{0}=N+1$ denotes the initial center position of the wave packet. In
Figs. \ref{fig2}(a)-(c), the trajectories $n_{a(b)}(t)$ are plotted as blue
(green) dashed lines, respectively, on top of the numerically obtained
probability distributions in the coordinate space.

In the regimes with a single effective ladder [Figs.\ \ref{fig2}(a) and
(b)], the dynamics is characterized by a single sinusoidal Bloch oscillation
with a well-defined period. By contrast, in the two-ladder regime [Fig.\ \ref%
{fig2}(c)], the interference between the two inequivalent ladders gives rise
to a pronounced beating envelope, which serves as a directly observable
hallmark of Bloch--Zener dynamics.

In summary, the quantum dynamics in the large-photon-number regime is
governed by the interplay between Wannier--Stark localization induced by the
cavity frequency and \textrm{SSH}-type dimerization originating from the
counter-rotating light--matter coupling. This interplay enables flexible and
robust control of Bloch and Bloch--Zener oscillations through experimentally
accessible parameters.

\section{Small-photon-number regime and localization induced by hopping
deformation}

\label{small_photon}

For the numerics in this section, we take the average photon number $\bar{n}%
=10$ as a representative small-photon-number value. In this regime, the
large-$\bar{n}$ approximation of the previous section is no longer
controlled. The photon-number dependence of the hopping matrix elements is
no longer weak enough for a perturbative treatment, and the matrix elements
associated with counter-rotating processes depend sensitively on the photon
number. As a result, the effective lattice in excitation-number space
becomes strongly inhomogeneous and deviates qualitatively from the large-$%
\bar{n}$ limit. The spectrum is therefore no longer organized by a single
simple ladder structure, and the dynamics becomes strongly state-dependent.
In particular, the well-defined Bloch-type motion supported by
well-separated Wannier--Stark ladders in the large-$\bar{n}$ regime is
replaced by transport governed by the detailed structure of the spectrum,
the density of states, and the spatial organization of the eigenstates.
\begin{figure*}[tbp]
\centering
\includegraphics[bb=3 398 1540 568, width=18cm, clip]{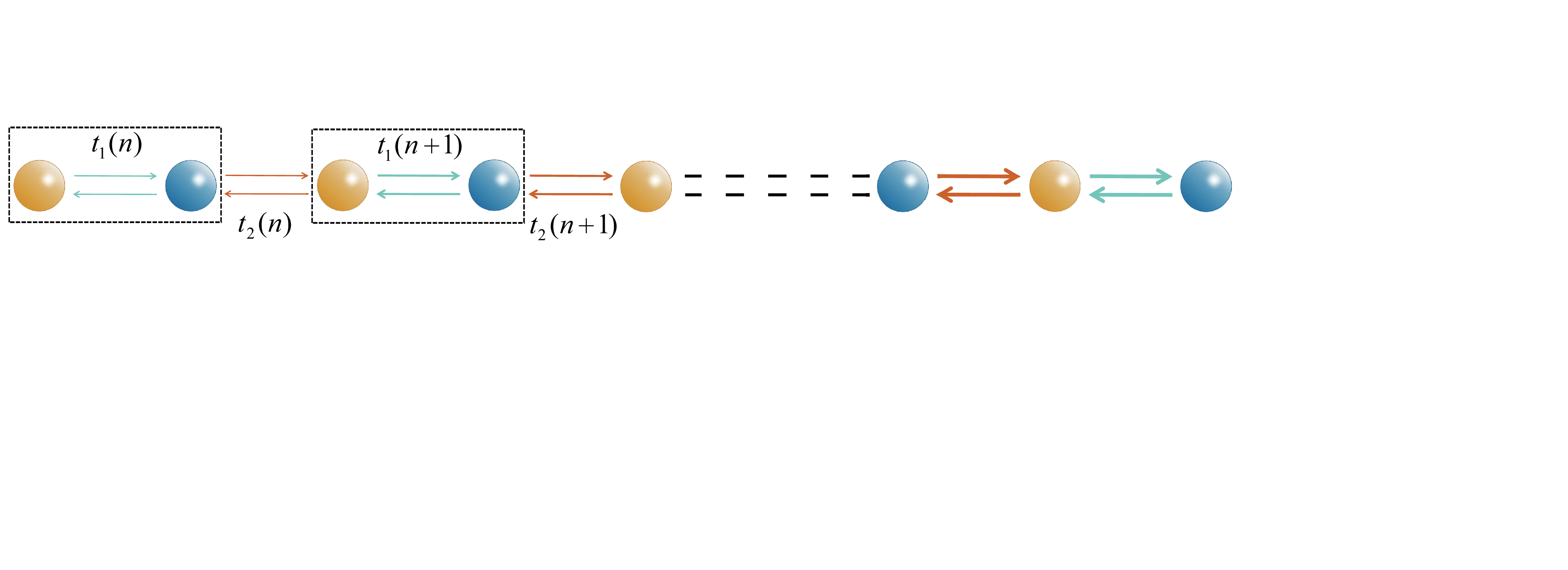}
\caption{Schematic of the effective lattice generated by $H_{\mathrm{eff},2}$
in the small-photon-number regime. Each unit cell (black dashed box)
consists of two sublattice sites coupled by the intracell hopping $t_{1}(n)=%
\protect\lambda _{2}\protect\sqrt{\bar{n}+2n-1}$, while adjacent cells are
connected by the intercell hopping $t_{2}(n)=\protect\lambda _{1}\protect%
\sqrt{\bar{n}+2n}$. Both couplings inherit the SSH-type dimerization ($%
t_{1}\neq t_{2}$) and grow monotonically with the cell index $n$, as
indicated by the increasing arrow thickness. The resulting excitation-number
lattice is therefore a position-dependent \textrm{SSH} chain rather than a
uniform dimerized lattice.}
\label{fig3}
\end{figure*}

\subsection{Hopping-deformed effective lattice and local-band picture}

\label{subsec:small_localband}

When $\bar{n}$ is no longer asymptotically large, the relations in Eq. (\ref%
{large condi}) cease to hold. Here we consider the even-parity case as an
example. The effective Hamiltonian in the odd-parity manifold therefore
takes the form
\begin{equation}
H_{\mathrm{eff},2}=H_{\mathrm{inc}}+H_{\omega }+H_{\Omega },  \label{Heff2}
\end{equation}%
with
\begin{equation}
H_{\mathrm{inc}}=\sum_{n=1}^{N}t_{1}(n)(|2n-1\rangle \langle
2n|)+t_{2}(n)(|2n\rangle \langle 2n+1|)+\mathrm{H.c.},  \label{Hinc}
\end{equation}%
where
\begin{equation}
t_{1}(n)=\lambda _{2}\sqrt{\bar{n}+2n-1},\text{ }t_{2}(n)=\lambda _{1}\sqrt{%
\bar{n}+2n}.  \label{eq:t1t2_small}
\end{equation}%
Equation~(\ref{Hinc}) defines a lattice with two intertwined ingredients.
First, the couplings are dimerized whenever $\lambda _{1}\neq \lambda _{2}$,
so the \textrm{SSH}-type intracell and intercell distinction is retained
(seen in Fig. \ref{fig3}). Second, both couplings increase monotonically
with the cell index $n$, which reflects the intrinsic photon-number
dependence of the transition amplitudes. This monotonic hopping gradient is
absent in the large-$\bar{n}$ limit and constitutes the essential new
feature of the small-photon-number regime.

A useful semi-analytic description is obtained by viewing Eq.~(\ref{Heff2})
as a local \textrm{SSH} lattice whose parameters vary with position.
Introducing the local cell basis
\begin{equation}
|A_{n}\rangle \equiv |2n-1\rangle ,\ |B_{n}\rangle \equiv |2n\rangle ,
\label{basis}
\end{equation}%
we define the local Bloch Hamiltonian
\begin{equation}
h_{n}(k)=\left(
\begin{array}{cc}
V_{A}(n) & t_{1}(n)+t_{2}(n)e^{-ik} \\
t_{1}(n)+t_{2}(n)e^{ik} & V_{B}(n)%
\end{array}%
\right) ,  \label{eq:local_bloch_small}
\end{equation}%
where
\begin{equation}
V_{A}(n)=[2n-1-(N+1)]\omega +\frac{\Omega }{2},  \label{eq:VAB_small1}
\end{equation}%
and
\begin{equation}
V_{B}(n)=[2n-(N+1)]\omega -\frac{\Omega }{2}.  \label{eq:VAB_small2}
\end{equation}%
The corresponding local bands are
\begin{eqnarray}
E_{\pm }(n,k) &=&V_{c}(n)\pm \lbrack \Delta
_{z}(n)^{2}+t_{1}(n)^{2}+t_{2}(n)^{2}  \notag \\
&&+2t_{1}(n)t_{2}(n)\cos k]^{1/2},  \label{eq:local_band_small}
\end{eqnarray}%
with
\begin{equation}
V_{c}(n)=\frac{V_{A}(n)+V_{B}(n)}{2},\text{ }\Delta _{z}(n)=\frac{%
V_{A}(n)-V_{B}(n)}{2}.  \label{eq:VcDz_small}
\end{equation}%
The local band edges are therefore
\begin{equation}
E_{\mathrm{edge}}^{\nu ,\mu }(n)=V_{c}(n)+\nu \sqrt{\Delta _{z}(n)^{2}+[\mu
t_{1}(n)+t_{2}(n)]^{2}},  \label{eq:local_edges_small}
\end{equation}%
with $\nu ,\mu =\pm $. This local-band picture already captures the main
spectral consequences of hopping deformation. Since both $t_{1}(n)$ and $%
t_{2}(n)$ increase with $n$, the local bandwidth becomes progressively
larger toward the strong-hopping side. As a result, large-$|E|$ states can
only be supported in cells with sufficiently large local bandwidth, which
immediately implies an energy-resolved spatial bias of the eigenstates.

The same framework also yields a turning-point criterion. For a given energy
$E$, local propagation is allowed only when the local quasi-momentum is
real, namely when
\begin{eqnarray}
\chi _{n}(E) &\equiv &\{[E-V_{c}(n)]^{2}-\Delta
_{z}^{2}(n)-t_{1}^{2}(n)-t_{2}^{2}(n)\}  \notag \\
&&\times \lbrack 2t_{1}(n)t_{2}(n)]^{-1},  \label{eq:chi_l_small}
\end{eqnarray}%
with real $k$ requiring $\chi _{n}(E)\in \lbrack -1,1]$. Equivalently, a
local turning point is reached when
\begin{equation}
|\chi _{n}(E)|\text{ }=1.  \label{eq:turning_small}
\end{equation}%
Outside the allowed region, the local quasimomentum becomes imaginary and
the wavefunction decays exponentially. In this sense, the localization
observed in the small-photon regime is not disorder induced. Rather, it
arises from a deterministic turning-point structure generated by the
position dependence of the local \textrm{SSH} bands.

\begin{figure*}[tbp]
\centering
\includegraphics[bb=75 43 1749 532, width=18.5cm, clip]{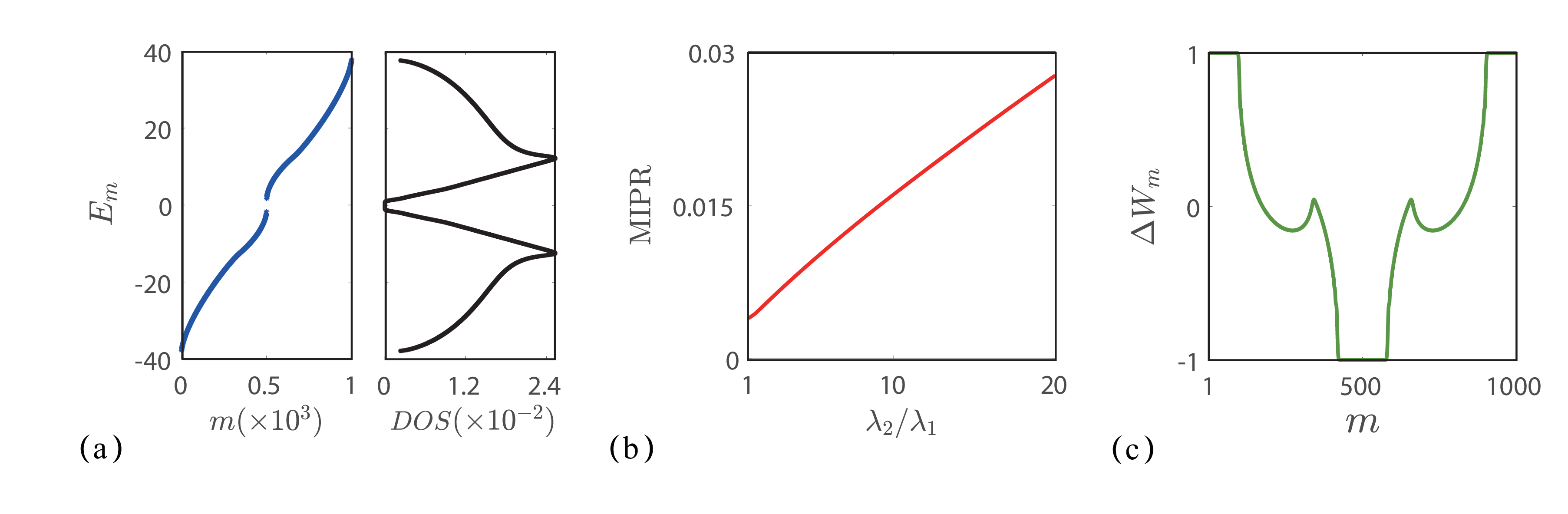}
\caption{Spectral structure and eigenstate diagnostics of the
hopping-deformed \textrm{SSH} model. (a) Eigenenergy structure $E_{m}$ (blue
curve) and density of states $\mathrm{DOS}$ (black curve) for representative
parameters. (b) Mean inverse participation ratio (\textrm{MIPR}) versus
dimerization ratio $\protect\lambda _{2}/\protect\lambda _{1}$: the
monotonic increase signals progressive localization as intra-cell hopping
becomes dominant over inter-cell coupling. (c) Right-left weight imbalance $%
\Delta W_{m}$ revealing an energy-dependent bias toward the strong-hopping
side. Parameters: $N=500$, $\bar{n}=10$, $\protect\lambda _{1}=0.4$ [panel
(b)] and $\protect\lambda _{2}=2\protect\lambda _{1}=0.8$ [panels (a),(c)].
Both the cavity frequency $\protect\omega $ and the atomic transition
frequency $\Omega $ are set to zero.}
\label{fig4}
\end{figure*}

\subsection{Spectral reconstruction and collective localization}

\label{subsec:small_spectrum}

To isolate the hopping deformation, we set $\omega =\Omega =0$ so that $H_{%
\mathrm{eff},2}=H_{\mathrm{inc}}$, and numerically diagonalize the
eigenvalue problem
\begin{equation}
H_{\mathrm{eff},2}|\Psi _{n}\rangle =H_{\mathrm{inc}}|\Psi _{n}\rangle
=E_{n}|\Psi _{n}\rangle .
\end{equation}%
A global characterization of the spectrum is provided by the density of
states (\textrm{DOS}), which we evaluate in the smoothed form
\begin{equation}
D(E)=\frac{1}{2N+1}\sum_{n=1}^{2N+1}g_{\eta }(E-E_{n}),  \label{DOS_def}
\end{equation}%
where $g_{\eta }(x)=(2\pi \eta ^{2})^{-1/2}e^{[-x^{2}/(2\eta ^{2})]}$ is a
normalized Gaussian broadening kernel of width $\eta $. Within the
local-band approximation, the same quantity can be interpreted as a
superposition of local contributions,
\begin{equation}
D(E)\approx \frac{1}{N}\sum_{n=1}^{N}D_{\mathrm{loc}}(E,n),
\label{eq:DOS_loc_sum}
\end{equation}%
with
\begin{equation}
D_{\mathrm{loc}}(E,n)=\frac{1}{2}\sum_{\nu =\pm }\int_{-\pi }^{\pi }\frac{dk%
}{2\pi }\delta \lbrack E-E_{\nu }(n,k)].  \label{eq:DOS_loc_def}
\end{equation}%
Equations (\ref{eq:local_band_small}) and (\ref{eq:DOS_loc_sum}) clarify why
the \textrm{DOS} is strongly reshaped by hopping deformation. Since the
local bandwidth increases with $n$, different energy windows receive
contributions from different spatial regions. This produces both spectral
inhomogeneity and energy-resolved spatial bias. As shown in Fig.\ \ref{fig4}%
(a), hopping deformation reshapes the spectrum and redistributes spectral
weight. Compared to the uniform-hopping case, the \textrm{DOS} is reduced
near the edges and enhanced around the spectral center, and the location of
the \textrm{DOS} maxima shifts as the dimerization is tuned. This reshaping
anticipates substantial changes in eigenstate structure and transport.

To quantify localization, we compute the inverse participation ratio (%
\textrm{IPR}) in the site basis $\{|n\rangle \}$,
\begin{equation}
\mathrm{IPR}(m)=\frac{\sum_{n=1}^{2N+1}|\langle n|\Psi _{m}\rangle |^{4}}{%
(\sum_{n=1}^{2N+1}|\langle n|\Psi _{m}\rangle |^{2})^{2}},
\label{eq:IPR_small}
\end{equation}%
and the mean inverse participation ratio
\begin{equation}
\mathrm{MIPR}=\frac{1}{2N+1}\sum_{m=1}^{2N+1}\mathrm{IPR}(m).
\label{eq:MIPR_small}
\end{equation}%
Figure \ref{fig4}(b) shows that $\mathrm{MIPR}$ increases from near zero in
the uniform limit to a finite value once hopping deformation is present.
This indicates that localization is enhanced for a broad portion of the
spectrum, rather than being a feature of isolated states. A finite \textrm{%
MIPR} indicates that a substantial fraction of the spectrum consists of
non-extended states, in agreement with the turning-point interpretation.

To reveal the spatial bias of the eigenstates, we further evaluate the
right-left weight imbalance
\begin{equation}
\Delta W_{m}=\sum_{n=N+2}^{2N+1}|\langle n|\Psi _{m}\rangle
|^{2}-\sum_{n=1}^{N+1}|\langle n|\Psi _{m}\rangle |^{2}.  \label{DeltaW_def}
\end{equation}

As shown in Fig. \ref{fig4}(c), $\Delta W_{m}$ exhibits an energy-dependent
bias toward the strong-hopping side, indicating that the hopping gradient
induces a deterministic, energy-resolved redistribution of spectral weight.
This observation lends further support to the above analysis. Qualitatively,
this behavior is distinct from the non-Hermitian skin effect \cite%
{WZ2018,Dora2025}, in which essentially all eigenstates accumulate at a
single boundary.

These quantities make explicit that the small-photon regime eigenstates are
not merely localized, but are in general biased toward the strong-hopping
side in an energy-dependent way. The origin of this bias is already encoded
in the local-band structure. Because the local bandwidth grows with $n$, the
low- and high-energy parts of the spectrum are preferentially supported by
cells on the strong-hopping side.

\begin{figure*}[tbp]
\centering
\includegraphics[bb=78 44 1787 535, width=18.5cm, clip]{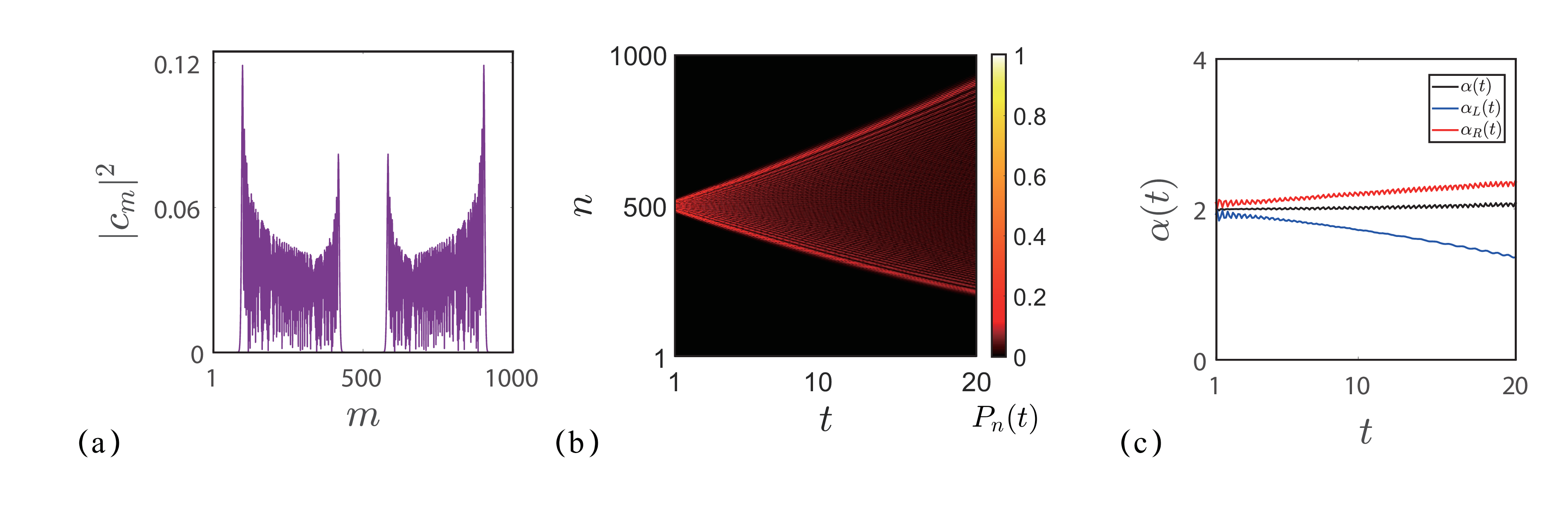}
\caption{Asymmetric diffusion in the small-photon-number regime. (a)
Spectral weights $|c_{m}|^{2}$ for the initial state $|\Phi (0)\rangle =%
\protect\delta _{n,N+1}|n\rangle $, the weighted imbalance $\overline{\Delta
W}=0.06>0$ indicating that the probability weight of initial state lies on
the strong-hopping side. (b) Time-evolution of the probability distribution $%
P_{n}(t)$ as defined in Eq. (\protect\ref{eq:psiPn_small}). (c)
Time-dependent diffusion exponents $\protect\alpha (t)$ (black solid line), $%
\protect\alpha _{L}(t)$ (blue solid line), and $\protect\alpha _{R}(t)$ (red
solid line) from Eqs. (\protect\ref{eq:alpha_t_small}) and (\protect\ref%
{eq:alphaLR_small}). The enhanced spreading toward the strong-hopping side
arises from the combined effect of large local currents and spatially biased
eigenstates. The global diffusion exponent $\protect\alpha \approx 2$
indicates direction-dependent transport with ballistic global spreading,
while the persistent oscillatory behavior and the ordering $\protect\alpha %
_{R}(t)>\protect\alpha (t)>\protect\alpha _{L}(t)$ directly reflect
interference from the dimerized band structure in conjunction with a
spatially increasing hopping amplitude. Parameters: $N=500$, $\bar{n}=10$,
and $\protect\lambda _{2}=2\protect\lambda _{1}=0.8$. Both the cavity
frequency $\protect\omega $ and the atomic transition frequency $\Omega $
are set to zero.}
\label{fig5}
\end{figure*}

\begin{figure*}[tbp]
\centering
\includegraphics[bb=61 32 2671 1604, width=20.5cm, clip]{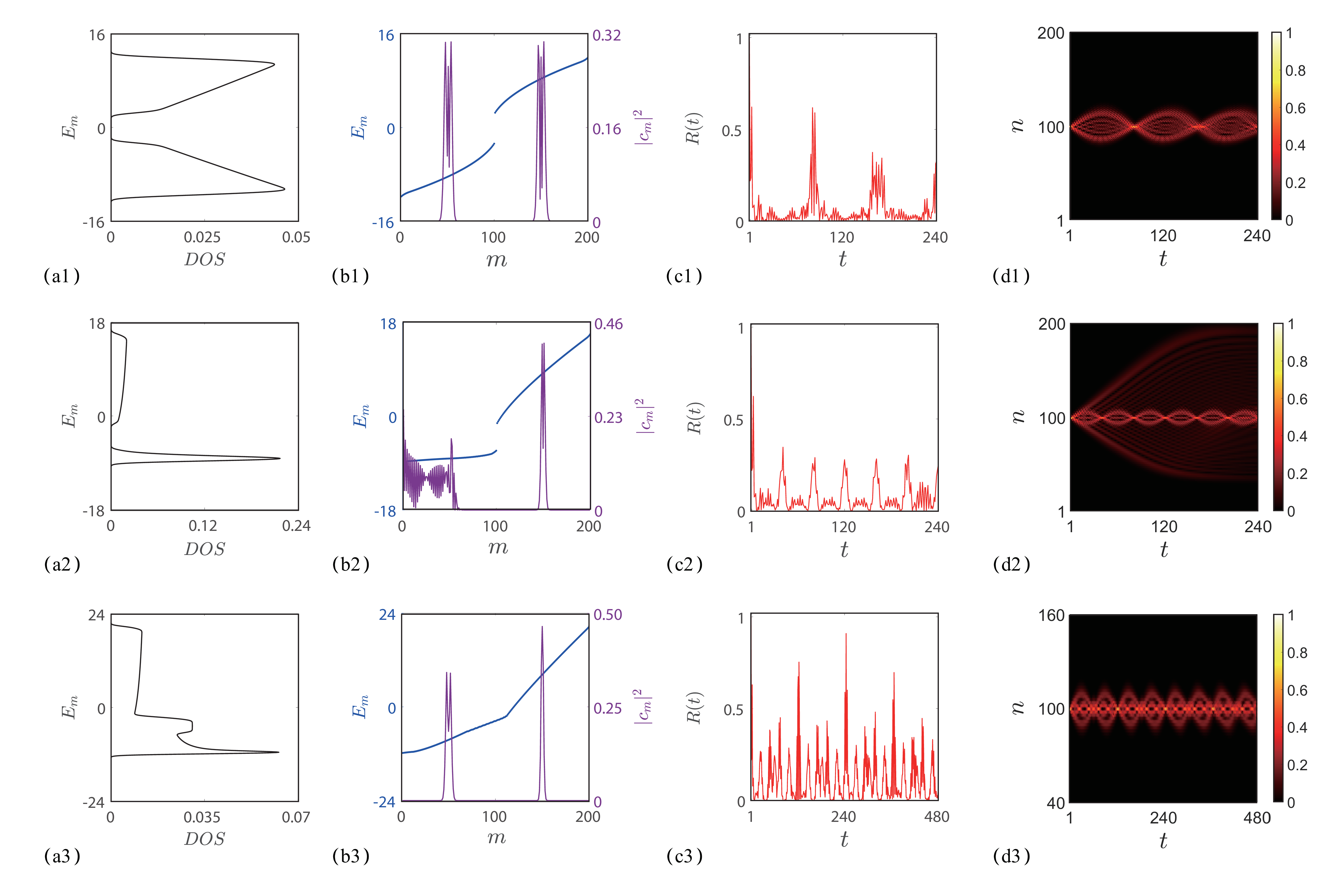}
\caption{Three dynamical regimes of the hopping-deformed \textrm{SSH}
lattice subjected to a Stark gradient. Each row corresponds to a different
Stark frequency, with $\protect\omega =0.001$, $\protect\omega =0.04$, $%
\protect\omega =0.09$ from top to bottom. The four columns present
complementary diagnostics: (a1-a3) density of states $\mathrm{DOS}$,
illustrating the spectral reshaping induced by hopping deformation; (b1-b3)
eigenenergy structure $E_{m}$ (blue, left axis) and spectral weights $%
|c_{m}|^{2}$ (purple, right axis) for the initial state $|\Phi (0)\rangle =%
\protect\delta _{n,N+1}|n\rangle $, highlighting the dominant energy windows
contributing to the dynamics; (c1-c3) return probability $R(t)$, exhibiting
beat-modulated revivals at small $\protect\omega $, partial revival with
dispersive broadening at intermediate $\protect\omega $, and multi-frequency
beating with multiple sharp revivals at large $\protect\omega $; (d1-d3)
time-evolution probability distribution $P_{n}(t)$, revealing beat-modulated
oscillations, coexistence of diffusion and oscillation, two rebuilt ladder
sectors, and multi-frequency beating. Parameters are $N=500$, $\bar{n}=10$, $%
\protect\lambda _{2}=0.8$, $\protect\lambda _{1}=0.03$. The resonance
condition $\Omega =\protect\omega $ is satisfied throughout.}
\label{fig6}
\end{figure*}

\subsection{Direction-dependent asymmetric diffusion and local kinetic
control}

\label{subsec_small_dyn}

We now examine the real-time dynamics generated by the same hopping
deformation. Starting from an initial state $|\Phi (0)\rangle $, we define
\begin{equation}
|\Phi (t)\rangle =e^{-iH_{\mathrm{eff},2}t}|\Phi (0)\rangle ,
\label{eq:time_small}
\end{equation}%
with site amplitudes
\begin{equation}
\psi _{n}(t)=\langle n|\Phi (t)\rangle ,\text{ }P_{n}(t)=|\psi _{n}(t)|^{2}.
\label{eq:psiPn_small}
\end{equation}%
The mean-square displacement (\textrm{MSD}) is
\begin{equation}
\sigma ^{2}(t)=\sum_{n=1}^{2N+1}(n-n_{0})^{2}P_{n}(t),  \label{MSD}
\end{equation}%
where $n_{0}$ is the center of the initial packet. To characterize spreading
we use the time-dependent diffusion exponent
\begin{equation}
\alpha (t)=\frac{d\ln \sigma ^{2}(t)}{d\ln t}.  \label{eq:alpha_t_small}
\end{equation}%
We also resolve the spreading into left and right parts,
\begin{equation}
\sigma _{L}^{2}(t)=\sum_{n=1}^{N+1}(n-n_{0})^{2}P_{n}(t),\text{ }\sigma
_{R}^{2}(t)=\sum_{n=N+2}^{2N+1}(n-n_{0})^{2}P_{n}(t),  \label{MSD_LR}
\end{equation}%
and define
\begin{equation}
\alpha _{L}(t)=\frac{d\ln \sigma _{L}^{2}(t)}{d\ln t},\text{ }\alpha _{R}(t)=%
\frac{d\ln \sigma _{R}^{2}(t)}{d\ln t}.  \label{eq:alphaLR_small}
\end{equation}%
The left-right asymmetry of the diffusion is not a consequence of any
asymmetry in the global spectrum. Instead, it follows from the interplay
between the energy-resolved spatial bias of the eigenstates and the position
dependence of the local kinetic scale.

To see this, we first write the Schr\"{o}dinger equation in the site
representation,
\begin{equation}
i\dot{\psi}_{n}(t)=J_{n-1}\psi _{n-1}(t)+V_{n}\psi _{n}(t)+J_{n}\psi
_{n+1}(t),  \label{eq:site_sch_small}
\end{equation}%
where $J_{n}=\langle n|H_{\mathrm{inc}}|n+1\rangle $ and $V_{n}=\langle n|H_{%
\mathrm{inc}}|n\rangle $. The corresponding discrete continuity equation is
\begin{equation}
\frac{d}{dt}P_{n}(t)=j_{n}(t)-j_{n-1}(t),  \label{continuity}
\end{equation}%
with probability current
\begin{equation}
j_{n}(t)=2J_{n}\mathrm{Im}\left[ \psi _{n}^{\ast }(t)\psi _{n+1}(t)\right] .
\label{current_def}
\end{equation}%
This immediately yields the bound
\begin{equation}
|j_{n}(t)|\leq 2J_{n}|\psi _{n}(t)||\psi _{n+1}(t)|.  \label{current_bound}
\end{equation}%
Equation (\ref{current_bound}) shows that the local hopping $J_{n}$ sets the
maximal local propagation speed. Since $J_{n}$ increases toward the
strong-hopping side, the same side also supports a larger local current.
This is the direct dynamical reason why the packet spreads faster there.

The second ingredient is the spectral decomposition
\begin{equation}
|\Phi (t)\rangle =\sum_{m}c_{m}e^{-iE_{m}t}|\Psi _{m}\rangle ,\text{ }%
c_{m}=\langle \Psi _{m}|\Phi (0)\rangle .  \label{eq:spectral_small}
\end{equation}%
Even when $|c_{m}|^{2}$ is approximately symmetric in energy, the excited
eigenstates can still carry a net spatial bias because $\Delta W_{m}$ is
energy dependent. A compact summary of this effect is the weighted imbalance
\begin{equation}
\overline{\Delta W}=\sum_{m}|c_{m}|^{2}\Delta W_{m}.  \label{DeltaW_weighted}
\end{equation}%
When $\overline{\Delta W}>0$, the initial state predominantly excites
eigenstates whose probability weight lies on the strong-hopping side, and
vice versa for $\overline{\Delta W}<0$.

In Fig. \ref{fig5}, the initial state is prepared as a single-site state $%
|\Phi (0)\rangle =\delta _{n,N+1}|n\rangle $ localized at the center of the
chain ($n=N+1$). Figure \ref{fig5}(a) shows the spectral weights $%
|c_{m}|^{2} $ of this initial state projected onto the eigenbasis of $H_{%
\mathrm{eff},2}$; this symmetric initial condition eliminates any bias
introduced by the choice of initial position. Figure \ref{fig5}(b) presents
the corresponding real-space dynamics, where the probability distribution $%
P_{n}(t)$ exhibits a pronounced asymmetric spreading. Figure \ref{fig5}(c)
shows that the global diffusion exponent saturates at $\alpha (t)\approx 2$,
indicating ballistic transport in the overall dynamics; however, the
ordering $\alpha _{R}(t)>\alpha (t)>\alpha _{L}(t)$ reveals that the
ballistic growth is driven almost entirely by the strong-hopping region,
while the weak-hopping side exhibits sub-ballistic oscillatory behavior. The
observed direction-dependent asymmetric diffusion is therefore the combined
consequence of two facts: the excited eigenstates are spatially biased
toward the strong-hopping side, and that same side supports a larger local
current because its bond strengths are larger.

\subsection{Competition between hopping deformation and the Stark gradient}

\label{subsec_small_fig6}

We finally demonstrate the three dynamical sectors displayed in Fig. \ref%
{fig6}. The key observation is that the on-site gradient and the hopping
deformation enter the local-band structure in qualitatively different ways.
Under the resonant condition $\Omega =\omega $, Eq. (\ref{eq:VcDz_small})
gives
\begin{equation}
V_{A}(n)=V_{B}(n)\equiv V_{c}(n),  \label{eq:resonant_equal_small}
\end{equation}%
with%
\begin{equation}
V_{c}(n)=(2n-N-\frac{3}{2})\omega ,\text{ }\Delta _{z}(n)=0
\label{eq:Vc_res_small}
\end{equation}%
The local bands then reduce to
\begin{equation}
E_{\pm }(n,k)=V_{c}(n)\pm \varepsilon _{n}(k),
\label{eq:local_band_res_small}
\end{equation}%
where
\begin{equation}
\varepsilon _{n}(k)=\sqrt{t_{1}(n)^{2}+t_{2}(n)^{2}+2t_{1}(n)t_{2}(n)\cos k}.
\label{eq:eps_l_small}
\end{equation}%
To understand the competition, it is useful to define
\begin{equation}
S(n)\equiv t_{1}(n)+t_{2}(n).  \label{eq:S_l_small}
\end{equation}%
The outer band edges can then be estimated as
\begin{equation}
E_{\mathrm{high}}(n)\approx V_{c}(n)+S(n),\text{ }E_{\mathrm{low}}(n)\approx
V_{c}(n)-S(n).  \label{eq:EhighElow_small}
\end{equation}%
Since both $V_{c}(n)$ and $S(n)$ increase with $n$, the high-energy branch
is shaped by two cooperating tendencies, whereas the low-energy branch is
shaped by two competing tendencies. Indeed,
\begin{equation}
\frac{dE_{\mathrm{high}}}{dn}\approx 2\omega +\frac{dS}{dn},\text{\ }\frac{%
dE_{\mathrm{low}}}{dn}\approx 2\omega -\frac{dS}{dn}.  \label{eq:dEdl_small}
\end{equation}%
The high-energy sector therefore remains monotonic and ladder-like more
easily, while the low-energy sector is destabilized first as $\omega $
increases. This is the fundamental reason why the low-energy quasi-ladder is
the first one to be destroyed in the intermediate-$\omega $ regime.

To connect this structure to the dynamics, we divide the spectrum into low-
and high-energy sectors, denoted by $S_{\mathrm{low}}$ and $S_{\mathrm{high}%
} $, and define their average level spacings
\begin{eqnarray}
\Delta E_{\mathrm{low}} &=&\left\langle E_{m+1}-E_{m}\right\rangle _{m\in S_{%
\mathrm{low}}},  \notag \\
\Delta E_{\mathrm{high}} &=&\left\langle E_{m+1}-E_{m}\right\rangle _{m\in
S_{\mathrm{high}}},  \label{eq:DeltaE_lowhigh_small}
\end{eqnarray}%
as well as the spectral weights carried by the chosen initial state,
\begin{equation}
W_{\mathrm{low}}=\sum_{m\in S_{\mathrm{low}}}|c_{m}|^{2},\text{ }W_{\mathrm{%
high}}=\sum_{m\in S_{\mathrm{high}}}|c_{m}|^{2}.  \label{eq:WlowWhigh_small}
\end{equation}%
To quantify the oscillatory component of the dynamics, we also monitor the
return probability
\begin{equation}
R(t)=|\langle \Phi (0)|\Phi (t)\rangle |^{2}.  \label{eq:return_small}
\end{equation}

The three dynamical regimes are then understood as follows.

\paragraph*{(i) Small $\protect\omega$: beat-modulated oscillations.}

When $\omega $ is weak, the hopping deformation dominates and both the
low-energy and high-energy sectors retain quasi-ladder structures with
comparable spacings,
\begin{equation}
\Delta E_{\mathrm{low}}\approx \Delta E_{\mathrm{high}}.
\end{equation}%
If the initial state has non-negligible overlap with both sectors, the
dynamics is governed by two nearby periods,
\begin{equation}
T_{\mathrm{low}}=\frac{2\pi }{\Delta E_{\mathrm{low}}},\text{ }T_{\mathrm{%
high}}=\frac{2\pi }{\Delta E_{\mathrm{high}}},
\end{equation}%
whose interference produces a slow beat envelope. This is the origin of the
beat-modulated Bloch-like oscillations in Fig. \ref{fig6}(a1-d1).

\paragraph*{(ii) Intermediate $\protect\omega$: coexistence of diffusion and
oscillation.}

As $\omega $ increases, the low-energy sector is destabilized first, because
Eq.\ (\ref{eq:dEdl_small}) shows that $E_{\mathrm{low}}(n)$ is controlled by
a competition between the linear gradient and the growing local bandwidth.
Once the low-energy quasi-ladder is destroyed, states in that sector become
dispersive and contribute primarily to direction-dependent anomalous
diffusion. By contrast, the high-energy sector is reinforced by the same
increase of $\omega $, since the two contributions in $E_{\mathrm{high}}(n)$
act cooperatively. As a consequence, if the initial state overlaps with both
sectors, part of the wave packet undergoes dispersive spreading while
another part displays BO-like revivals. This coexistence is precisely the
physical content of Fig.\ \ref{fig6}(a2-d2).

\paragraph*{(iii) Large $\protect\omega$: two rebuilt ladder sectors and
multi-frequency beating.}

When $\omega $ becomes sufficiently large, the Stark gradient dominates both
sectors. The low-energy sector, which was destabilized first, is now
reorganized into a new ladder-like structure, while the high-energy sector
remains ladder-like. The two sectors therefore recover quasi-equidistant
organization, but with clearly different spacings,
\begin{equation}
\Delta E_{\mathrm{low}}\neq \Delta E_{\mathrm{high}}.
\end{equation}%
The dynamics is again oscillatory, but now the two characteristic periods
are well separated, producing a qualitatively different beat pattern from
the small-$\omega $ regime. This is the origin of Fig.\ \ref{fig6}(a3-d3).

The small-photon-number regime therefore realizes a hopping-deformed \textrm{%
SSH}-like lattice in which spectral reshaping, collective localization, and
transport are tightly linked. The local-band picture explains the \textrm{DOS%
} reconstruction, the emergence of turning-point-induced localization, the
energy-resolved spatial bias of the eigenstates, and the direction-dependent
anomalous diffusion. Once the Stark gradient is included, the same framework
also explains why different spectral sectors respond differently to $\omega $
and why the dynamics organizes into the three regimes shown in Fig. \ref%
{fig6}.

\section{Summary and discussion}

\label{Summary}

We have studied a cavity--atom system beyond the rotating-wave approximation
by formulating its dynamics as an effective lattice problem in
excitation-number space. From the viewpoint of superconducting-circuit
quantum simulation, the model provides a concrete setting in which
counter-rotating light--matter processes are converted into tunable hopping
structures, ladder spectra, localization, and nontrivial transport. The
central result is that the same microscopic platform supports two
qualitatively distinct regimes controlled by the photon number.

In the large-photon-number regime, the photon-number dependence of the
matrix elements becomes weak and the effective lattice reduces to a
dimerized SSH-type chain in a linear energy gradient. The spectrum then
organizes into one or two Wannier--Stark ladders, and the resulting Bloch
and Bloch--Zener oscillations account quantitatively for the observed
revival patterns. In the small-photon-number regime, by contrast, the matrix
elements retain a strong photon-number dependence and generate a
hopping-deformed lattice with dimerized couplings that increase along the
chain. This deformation reshapes the density of states, enhances collective
localization, and produces an energy-resolved spatial bias of the
eigenstates. The corresponding real-time dynamics displays asymmetric
anomalous diffusion, Bloch-like revivals, and beat phenomena that can all be
traced back to the same local-band organization.

Taken together, these results show that counter-rotating coupling is not
merely a correction to resonant cavity dynamics. In the excitation-number
basis it becomes a mechanism for engineering synthetic transport landscapes
with experimentally accessible signatures. This identifies superconducting
cavity--atom architectures as a promising platform for exploring emergent
ladder physics, hopping-induced localization, and controlled nonequilibrium
dynamics within a single light--matter setting.

\acknowledgments This work was supported by the Young Scientific and
Technological Talents (Level Three) in Tianjin (Grant No. QN20250303), and
the National Natural Science Foundation of China (Grants No. 12305026,
12675024, 12275193).

\section*{DATA AVAILABILITY}

All data generated in this work are obtained from numerical simulations
based on the described model and are reproducible from the provided
equations and parameters. Source code is available from the corresponding
author upon reasonable request.

\appendix

\section{Detailed Mapping of Parity Subspaces}

\label{Appendix:A}

The Hilbert space of the system is spanned by the basis states $\{|g,
n\rangle, |e, n\rangle\}$, where $|g\rangle$ ($|e\rangle$) denotes the
atomic ground (excited) state and $|n\rangle$ is the photon number state. To
map the original Hamiltonian into a 1D lattice, we decompose the space into
two decoupled subspaces based on the parity of the total excitation number.

The odd-excitation subspace $H_{o}$ is defined by states with an odd number
of total excitations. We map these states to a 1D site basis $\{|n\rangle
_{o}\}$ as follows:
\begin{equation}
|e,\bar{n}+2n-2\rangle \rightarrow |2n-1\rangle _{o},\quad |g,\bar{n}%
+2n-1\rangle \rightarrow |2n\rangle _{o}.
\end{equation}%
The actions of the Hamiltonian $H$ on these basis states are:
\begin{align}
H|2n\rangle _{o}& =[\omega (\bar{n}+2n-1)-\frac{\Omega }{2}]|2n\rangle _{o}
\notag \\
& +\lambda _{1}\sqrt{\bar{n}+2n-1}|2n-1\rangle _{o}  \notag \\
& +\lambda _{2}\sqrt{\bar{n}+2n}|2n+1\rangle _{o}
\end{align}%
and
\begin{eqnarray}
H|2n+1\rangle _{o} &=&[\omega (\bar{n}+2n)+\frac{\Omega }{2}]|2n+1\rangle
_{o}  \notag \\
&&+\lambda _{1}\sqrt{\bar{n}+2n+1}|2n+2\rangle _{o}  \notag \\
&&+\lambda _{2}\sqrt{\bar{n}+2n}|2n\rangle _{o}.
\end{eqnarray}

Similarly, the even-excitation subspace $H_{e}$\ follows the mapping:
\begin{equation}
|g,\bar{n}+2n-2\rangle \rightarrow |2n-1\rangle _{e},\quad |e,\bar{n}%
+2n-1\rangle \rightarrow |2n\rangle _{e}.
\end{equation}%
The corresponding Hamiltonian actions are:
\begin{align}
H|2n\rangle _{e}& =[\omega (\bar{n}+2n-1)+\frac{\Omega }{2}]|2n\rangle _{e}
\notag \\
& +\lambda _{1}\sqrt{\bar{n}+2n}|2n+1\rangle _{e}  \notag \\
& +\lambda _{2}\sqrt{\bar{n}+2n-1}|2n-1\rangle _{e}
\end{align}%
and
\begin{eqnarray}
H|2n+1\rangle _{e} &=&[\omega (\bar{n}+2n)-\frac{\Omega }{2}]|2n+1\rangle
_{e}  \notag \\
&&+\lambda _{1}\sqrt{\bar{n}+2n}|2n\rangle _{e}  \notag \\
&&+\lambda _{2}\sqrt{\bar{n}+2n+1}|2n+2\rangle _{e}.
\end{eqnarray}

Note that for the even subspace, the dimerized hopping parameters $\lambda
_{1}$ and $\lambda _{2}$ are effectively swapped relative to the odd
subspace.

\section{Detailed Perturbative Derivation}

\label{Appendix:B}

We consider the Hamiltonian in the even-parity subspace, where the
unperturbed eigenstates are given by $|\varphi _{m}\rangle
=\sum_{n}J_{n-m}(-2\kappa /\omega )|n\rangle $ with unperturbed energies $%
\varepsilon _{m}=[m-(N+1)]\omega $. The first-order energy correction is $%
\varepsilon _{m}^{(1)}=\langle \varphi _{m}|H_{\Delta }+H_{\Omega }|\varphi
_{m}\rangle $. For the dimerization term $H_{\Delta }$, substituting the
expansion of $|\varphi _{m}\rangle $:
\begin{eqnarray}
\langle \varphi _{m}|H_{\Delta }|\varphi _{m}\rangle &=&\kappa \Delta
\sum_{n}[J_{2n-1-m}(z)J_{2n-m}(z)  \notag \\
&&+J_{2n-m}(z)J_{2n-1-m}(z)-J_{2n-m}J_{2n+1-m}(z)  \notag \\
&&-J_{2n+1-m}J_{2n-m}(z)],
\end{eqnarray}%
with the Bessel function argument $z=-2\kappa /\omega $. Simplifying the
terms and using the derivative identity $J_{2n-1-m}(z)-J_{2n+1-m}(z)=2\frac{d%
}{dz}J_{2n-m}(z)$, we obtain:
\begin{eqnarray}
\langle \varphi _{m}|H_{\Delta }|\varphi _{m}\rangle &=&4\kappa \Delta
\sum_{n}J_{2n-m}(z)\frac{d}{dz}J_{2n-m}(z)  \notag \\
&=&2\kappa \Delta \frac{d}{dz}\sum_{n}[J_{2n-m}(z)]^{2}.
\end{eqnarray}%
Applying the summation formulas from Appendix \ref{Appendix:C}, we have
\begin{eqnarray}
\langle \varphi _{m}|H_{\Delta }|\varphi _{m}\rangle &=&2\kappa \Delta \frac{%
d}{dz}\{\frac{1}{2}[1+(-1)^{m}J_{0}(2z)]\}  \notag \\
&=&-2\kappa \Delta (-1)^{m}J_{1}(2z).
\end{eqnarray}%
For the potential term $H_{\Omega }$, we have
\begin{equation}
\langle \varphi _{m}|H_{\Omega }|\varphi _{m}\rangle =\frac{\Omega }{2}%
\sum_{n}(-1)^{n}[J_{n-m}(z)]^{2}=\frac{\Omega }{2}(-1)^{m}J_{0}(2z).
\end{equation}%
The corrected energy
\begin{equation}
E_{m}=\varepsilon _{m}+(-1)^{m}[\frac{\Omega }{2}J_{0}(2z)-2\kappa \Delta
J_{1}(2z)].
\end{equation}

The first-order state correction is $|\varphi _{m}^{(1)}\rangle =\sum_{n\neq
m}\frac{\langle \varphi _{n}|V|\varphi _{m}\rangle }{\varepsilon
_{m}-\varepsilon _{n}}|\varphi _{n}\rangle $, where $V=H_{\Delta }+H_{\Omega
}$. In the large-$\omega $ limit, the terms with $n=m\pm 1$ dominate. For
the dimerization part:
\begin{equation}
\frac{\langle \varphi _{m-1}|H_{\Delta }|\varphi _{m}\rangle }{\omega }=%
\frac{\kappa \Delta (-1)^{m}}{\omega }[J_{0}(2z)-J_{2}(2z)].
\end{equation}%
For the potential part:
\begin{equation}
\frac{\langle \varphi _{m-1}|H_{\Omega }|\varphi _{m}\rangle }{\omega }=%
\frac{\Omega }{2\omega }(-1)^{m}J_{1}(2z).
\end{equation}%
Combining these, the corrected state $|\Psi _{m}\rangle =|\varphi
_{m}\rangle +|\varphi _{m}^{(1)}\rangle $ matches the expression in the main
text:
\begin{eqnarray}
|\Psi _{m}\rangle &=&|\varphi _{m}\rangle +\frac{(-1)^{m}}{2\omega }%
\{2\kappa \Delta \lbrack J_{0}(2z)-J_{2}(2z)]  \notag \\
&&+\Omega J_{1}(2z)\}(|\varphi _{m-1}\rangle +|\varphi _{m+1}\rangle ).
\end{eqnarray}

\section{Bessel Function Identities}

\label{Appendix:C}

The derivations in Appendix \ref{Appendix:B} rely on the following
properties of Bessel functions of the first kind, $J_{n}(z)$:

1. Parity Relation:
\begin{equation}
J_{-v}(z)=(-1)^{v}J_{v}(z).
\end{equation}

2. Recurrence Relations and Derivatives:
\begin{equation}
\left\{
\begin{array}{c}
J_{\nu -1}(z)-J_{\nu +1}(z)=2\frac{d}{dz}J_{\nu }(z) \\
J_{\nu -1}(z)+J_{\nu +1}(z)=(2v/z)J_{\nu }(z)%
\end{array}%
\right. .
\end{equation}

3. Summation Formulas:
\begin{equation}
\left\{
\begin{array}{c}
\sum_{n=-\infty }^{\infty }J_{n}(z)J_{n+k}(z)=\delta _{k,0} \\
\sum_{n=-\infty }^{\infty }(-1)^{n}J_{n}(z)J_{n+k}(z)=J_{k}(2z) \\
\sum_{n=-\infty }^{\infty }\left[ J_{2n-m}(z)\right] ^{2}=\frac{1}{2}\left[
1+(-1)^{m}J_{0}(2z)\right]%
\end{array}%
\right. .
\end{equation}


\begin{thebibliography}{99}
\bibitem{Blais2021} A. Blais, A. L. Grimsmo, S. M. Girvin, and A. Wallraff,
Circuit quantum electrodynamics, Rev. Mod. Phys. 93, 025005 (2021).

\bibitem{JC1963} E. T. Jaynes and F. W. Cummings, Comparison of quantum and
semiclassical radiation theories with application to the beam maser, Proc.
IEEE \textbf{51}, 89 (1963).

\bibitem{Raimond2001} J. M. Raimond, M. Brune, and S. Haroche, Manipulating
quantum entanglement with atoms and photons in a cavity, Rev. Mod. Phys.
\textbf{73}, 565 (2001).

\bibitem{Thompson1992} R. J. Thompson, G. Rempe, and H. J. Kimble,
Observation of normal-mode splitting for an atom in an optical cavity, Phys.
Rev. Lett. \textbf{68}, 1132 (1992).

\bibitem{Ritsch2021} F. Mivehvar, F. Piazza, T. Donner, and H. Ritsch,
Cavity QED with quantum gases: new paradigms in many-body physics, Adv.
Phys. \textbf{70}, 1 (2021).

\bibitem{Andreas2022} A. Reiserer, Colloquium: Cavity-enhanced quantum
network nodes, Rev. Mod. Phys. \textbf{94}, 041003 (2022).

\bibitem{David2022} S. Chakram, K. He, A. V. Dixit, A. E. Oriani, R. K.
Naik, N. Leung, H. Kwon, W.-L. Ma, L. Jiang, and D. I. Schuster, Multimode
photon blockade, Nat. Phys. \textbf{18}, 879 (2022).

\bibitem{Wallraff2004} A. Wallraff, D. I. Schuster, A. Blais, L. Frunzio,
R.-S. Huang, J. Majer, S. Kumar, S. M. Girvin, and R. J. Schoelkopf, Strong
coupling of a single photon to a superconducting qubit using circuit quantum
electrodynamics, Nature \textbf{431}, 162 (2004).

\bibitem{Simmonds2014} M. S. Allman, J. D. Whittaker, M.
Castellanos-Beltran, K. Cicak, F. da Silva, M. P. DeFeo, F. Lecocq, A.
Sirois, J. D. Teufel, J. Aumentado, and R. W. Simmonds, Tunable resonant and
nonresonant interactions between a phase qubit and LC resonator, Phys. Rev.
Lett. \textbf{112}, 123601 (2014).

\bibitem{Solano2019} P. Forn-D\'{\i}az, L. Lamata, E. Rico, J. Kono, and E.
Solano, Ultrastrong coupling regimes of light-matter interaction, Rev. Mod.
Phys. \textbf{91}, 025005 (2019).

\bibitem{Nori2024} W. Qin, A. F. Kockum, C. S\'{a}nchez Mu\~{n}oz, A.
Miranowicz, and F. Nori, Quantum amplification and simulation of strong and
ultrastrong coupling of light and matter, Phys. Rep. \textbf{1078}, 1 (2024).

\bibitem{Alejandro2024} A. Gonz\'{a}lez-Tudela, A. Reiserer, J. J. Garc\'{\i}%
a-Ripoll, and F. J. Garc\'{\i}a-Vidal, Light-matter interactions in quantum
nanophotonic devices, Nat. Rev. Phys. \textbf{6}, 166 (2024).

\bibitem{Celi2014} A. Celi, P. Massignan, J. Ruseckas, N. Goldman, I. B.
Spielman, G. Juzeli\={u}nas, and M. Lewenstein, Synthetic gauge fields in
synthetic dimensions, Phys. Rev. Lett. \textbf{112}, 043001 (2014).

\bibitem{Goldman2016} N. Goldman, J. C. Budich, and P. Zoller, Topological
quantum matter with ultracold gases in optical lattices, Nat. Phys. \textbf{%
12}, 639 (2016).

\bibitem{Tomoki2019} T. Ozawa and H. M. Price, Topological quantum matter in
synthetic dimensions, Nat. Rev. Phys. \textbf{1}, 349 (2019).

\bibitem{Segev2021} E. Lustig and M. Segev, Topological photonics in
synthetic dimensions, Adv. Opt. Photonics \textbf{13}, 426 (2021).

\bibitem{GGC2022} M. Yang, J.-S. Xu, C.-F. Li, and G.-C. Guo, Simulating
topological materials with photonic synthetic dimensions in cavities,
Quantum Front. \textbf{1}, 10 (2022).

\bibitem{Nori2019} A. F. Kockum, A. Miranowicz, S. De Liberato, S. Savasta,
and F. Nori, Ultrastrong coupling between light and matter, Nat. Rev. Phys.
\textbf{1}, 19 (2019).

\bibitem{Niemczyk2010} T. Niemczyk, F. Deppe, H. Huebl, E. P. Menzel, F.
Hocke, M. J. Schwarz, J. J. Garc\'{\i}a-Ripoll, D. Zueco, T. H\"{u}mmer, E.
Solano, A. Marx, and R. Gross, Circuit quantum electrodynamics in the
ultrastrong-coupling regime, Nat. Phys. \textbf{6}, 772 (2010).

\bibitem{Rabi1937} I. I. Rabi, Space quantization in a gyrating magnetic
field, Phys. Rev. \textbf{51}, 652 (1937).

\bibitem{Braak2011} D. Braak, Integrability of the Rabi model, Phys. Rev.
Lett. \textbf{107}, 100401 (2011).

\bibitem{XQT2014} Q.-T. Xie, S. Cui, J.-P. Cao, L. Amico, and H. Fan,
Anisotropic Rabi model, Phys. Rev. X \textbf{4}, 021046 (2014).

\bibitem{Jochen2017} J. Braum\"{u}ller, M. Marthaler, A. Schneider, A.
Stehli, H. Rotzinger, M. Weides, and A. V. Ustinov, Analog quantum
simulation of the Rabi model in the ultra-strong coupling regime, Nat.
Commun. \textbf{8}, 779 (2017).

\bibitem{YL2017} Y. Lu, S. Chakram, N. Leung, N. Earnest, R. K. Naik, Z.
Huang, P. Groszkowski, E. Kapit, J. Koch, and D. I. Schuster, Universal
stabilization of a parametrically coupled qubit, Phys. Rev. Lett. \textbf{119%
}, 150502 (2017).

\bibitem{DiCarlo2017} N. K. Langford, R. Sagastizabal, M. Kounalakis, C.
Dickel, A. Bruno, F. Luthi, D. J. Thoen, A. Endo, and L. DiCarlo,
Experimentally simulating the dynamics of quantum light and matter at
deep-strong coupling, Nat. Commun. \textbf{8}, 1715 (2017).

\bibitem{CWW2019} W. Cai, J. Han, F. Mei, Y. Xu, Y. Ma, X. Li, H. Wang, Y.
P. Song, Z.-Y. Xue, Z.-Q. Yin, S. Jia, and L. Sun, Observation of
topological magnon insulator states in a superconducting circuit, Phys. Rev.
Lett. \textbf{123}, 080501 (2019).

\bibitem{GGC2021} X.-Y. Guo, Z.-Y. Ge, H. Li, Z. Wang, Y.-R. Zhang, P. Song,
Z. Xiang, X. Song, Y. Jin, L. Lu, K. Xu, D. Zheng, and H. Fan, Observation
of Bloch oscillations and Wannier--Stark localization on a superconducting
quantum processor, npj Quantum Inf. \textbf{7}, 51 (2021).

\bibitem{XQK2024} X. Li, H. Xu, J. Wang, L.-Z. Tang, D.-W. Zhang, C. Yang,
T. Su, C. Wang, Z. Mi, W. Sun, X. Liang, M. Chen, C. Li, Y. Zhang, K.
Linghu, J. Han, W. Liu, Y. Feng, P. Liu, G. Xue, et al., Mapping the
topology-localization phase diagram with quasiperiodic disorder using a
programmable superconducting simulator, Phys. Rev. Research \textbf{6},
L042038 (2024).

\bibitem{SPT2024} P. Song, Z. Xiang, Y.-X. Zhang, Z. Wang, X. Guo, X. Ruan,
X. Song, K. Xu, Y. Y. Gao, H. Fan, and D. Zheng, Coherent control of Bloch
oscillations in a superconducting circuit, PRX Quantum \textbf{5}, 020302
(2024).

\bibitem{ZJJ2025} J. Zhang, W. Huang, J. Chu, J. Qiu, X. Sun, Z. Tao, J.
Zhang, L. Zhang, Y. Zhou, Y. Chen, Y. Liu, S. Liu, Y. Zhong, J.-J. Miao, J.
Niu, and D. Yu, Synthetic multidimensional Aharonov-Bohm cages in Fock state
lattices, Phys. Rev. Lett. \textbf{134}, 070601 (2025).

\bibitem{Jonas2023} P. Saugmann and J. Larson, Fock-state-lattice approach
to quantum optics, Phys. Rev. A \textbf{108}, 033721 (2024).

\bibitem{Metzler1999} R. Metzler, E. Barkai, and J. Klafter, Anomalous
diffusion and relaxation close to thermal equilibrium: A fractional
Fokker-Planck equation approach, Phys. Rev. Lett. \textbf{82}, 3563 (1999).

\bibitem{ZH2006} H. Zhao, Identifying diffusion processes in one-dimensional
lattices in thermal equilibrium, Phys. Rev. Lett. \textbf{96}, 140602 (2006).

\bibitem{ZHP2024} H. P. Zhang, K. L. Zhang, and Z. Song, Dynamics of
non-Hermitian Floquet Wannier--Stark system, New J. Phys. \textbf{26},
123020 (2024).

\bibitem{WR2024} R. Wang and Z. Song, Flat band and $\eta $-pairing states
in a one-dimensional Moir\'{e} Hubbard model, Chin. Phys. Lett. \textbf{41},
047101 (2024).

\bibitem{ZHP2024-2} H. P. Zhang and Z. Song, Extended Wannier--Stark ladder
and electron-pair Bloch oscillations in dimerized non-Hermitian systems,
Phys. Rev. B \textbf{110}, 064310 (2024).

\bibitem{ZHP2025} H. P. Zhang and Z. Song, Dynamic Moir\'{e}-like pattern in
non-Hermitian Wannier--Stark ladder system, Ann. Phys. (N.Y.) \textbf{479},
170049 (2025).

\bibitem{Jonas2018} A. Gagge and J. Larson, Bloch-like energy oscillations,
Phys. Rev. A \textbf{98}, 053820 (2018).

\bibitem{ZG2015} G. Zhang, W. H. Hu, and Z. Song, Amplitude modulated Bloch
oscillations of photon probability distribution in a cavity-atom system, J.
Phys. B: At. Mol. Opt. Phys. \textbf{48}, 075402 (2015).

\bibitem{Maksimov2015} D. N. Maksimov, E. N. Bulgakov, and A. R. Kolovsky,
Wannier--Stark states in double-periodic lattices. I. One-dimensional
lattices, Phys. Rev. A \textbf{91}, 053631 (2015).

\bibitem{WZ2018} S. Y. Yao and Z. Wang, Edge States and Topological
Invariants of Non-Hermitian Systems, Phys. Rev. Lett. \textbf{121}, 086803
(2018).

\bibitem{Dora2025} B. Het\'{e}nyi and B. D\'{o}ra, Localized states and skin
effect around non-Hermitian impurities in tight-binding models, Phys. Rev. B
\textbf{112}, 075123 (2025).
\end{thebibliography}
\end{document}